\documentclass[reprint,amsmath,amssymb,aps,prl]{revtex4-1}
\usepackage[english] {babel}
\usepackage{hyperref}
\usepackage{amsfonts,graphicx}
\usepackage{hyperref}
\usepackage{subcaption}
\usepackage{epstopdf}

\begin{document}

\title{Field Electron Emission Induced Glow Discharge in Nanodiamond Vacuum Diode}

\author{Stanislav S. Baturin}
\email{s.s.baturin@gmail.com}
\affiliation{PSD Enrico Fermi Institute, The University of Chicago, 5640 S. Ellis Ave., Chicago, IL 60637, USA}
\author{Tanvi Nikhar} 
\author{Sergey V. Baryshev}
\email{serbar@msu.edu}
\affiliation{Department of Electrical and Computer Engineering, Michigan State University, 428 S. Shaw Ln., East Lansing, MI 48824, USA}
\date{\today}

\begin{abstract}
\noindent 
The present letter extends the prior findings on self-induced heating of solid state field emission devices. It was found that a vacuum diode (base pressure $\sim10^{-9}$ Torr), that makes use of graphite-rich polycrystalline diamond as cathode material, can switch from diode regime to resistor regime, to glow discharge plasma regime without any external perturbation, i.e. all transitions are self-induced. Combined results of \textit{in situ} field emission microscopy and \textit{ex situ} electron microscopy and Raman spectroscopy suggested that the nanodiamond cathode of the diode heated to about 3000 K which caused self-induced material evaporation, ionization and eventually micro-plasma formation. Our results confirm that field emission, commonly called cold emission, is a very complex phenomenon that can cause severe thermal load. Thermal load and material runaway could be the major factors causing vacuum diode deterioration, i.e. progressive increase in turn-on field, decrease in field enhancement factor, and eventual failure.
\end{abstract}

\maketitle 

Field emitters are ubiquitous electron sources found in microscopy, medical, mass spectrometry, and high power high energy machines and systems \cite{1,2,3,4,5,6}. However, utilization of FE sources is a trade-off between utmost performance (highest current) and lifetime. At high output current densities, obtained at very high local electric fields, field emitter may rapidly fail to operate. Failures are often accompanied with vacuum arcs (also called vacuum breakdowns) when plasma forms in high and ultrahigh vacuum environment. During arcs, surface of the emitter irreversibly modifies with the net mass loss through volatilization. Vacuum breakdown is studied with highest attention at CERN and SLAC in the effort to increase the accelerating gradient (measured in MV per m), the accelerator figure of merit \cite{7} and to minimize the operational downtime of high gradient linear accelerators. Before going into operation, accelerating structures are conditioned for weeks, i.e. they are exposed to many millions of short exceptionally high power electromagnetic pulses to purposely destroy undesired field emission (dark current) centers. During conditioning, many hundred thousand breakdowns take place leaving behind extensively dented and eroded surface \cite{8}.
In plasma-related disciplines, there is a common agreement that the field emission is the precursor and trigger of vacuum breakdowns and discharges. Many general correlations and scaling laws were established and verified phenomenologically. At the same time, the detailed transition from field electron emission to plasma formation is poorly understood and is under debate due to high complexity of the problem with many possible mechanisms involved. For example, an argument still remains whether such transitions are thermally driven or not. Norem $et$ $al.$ \cite{9} and Antoine $et$ $al.$ \cite{10} argued that the initial copper surface failure and a field emitting asperity breakup and volatilization is a cold process due to electromigration. In contrast, in the recent work by Kyritsakis $et$ $al.$ \cite{11} it was computationally found for a copper asperity that breakdown formation and plasma discharge are thermally activated via Nottingham heating/cooling and Joule heating channels.

Accelerating structures are relatively large volume standing and traveling wave high vacuum devices in which breakdowns are extremely short lived (on the nanosecond timescale) and therefore are extremely hard to study. However, arcs and breakdowns are ubiquitously present not only in accelerator R$\&$D but in many other disciplines \cite{12,13,14}, meaning that there are many other platforms that can provide insight into the vacuum breakdown problem. The vacuum diode, a configuration often utilized for electron source prototyping/testing, could be one of such platforms. Indeed, in recent years several reports appeared that discussed transition from field emission to stable long lasting plasma discharge regime in small volume microgap DC vacuum diodes \cite{15, 16}. When such transition takes place, it was found that the current vs. voltage relation switches from Fowler-Nordheim to DC gas breakdown/discharge dependence. The transition is also accompanied with visible light generation. It was proposed that the plasma formation was due to anode material evaporation, accumulation and ionization. This supports earlier findings on anodic plasma formation for sensitized carbon fiber cathodes operated in pulsed $\mu$s regime \cite{17} but contradicts findings for accelerator cavities in which only cathodic plasma is possible (the entire inner surface of the cavity is the cathode). It also contradicts very recent results obtained for a high voltage $\mu$s pulse vacuum copper tip-copper plate diode \cite{18}. Therefore, currently a second contradiction exists, whether the discharge is due to anodic or cathodic plasma formation.

In the present study we use a DC vacuum diode that makes use of nitrogen-incorporated highly conductive \textit{n}-type ultrananocrystalline diamond (UNCD) cathode as a model platform to test the hypothesis of the hot cathodic breakdown. Unlike the common vacuum diode design that relies on sharp nanoscopic tips, convenience of using UNCD as a cathode material stems from that it has large emission area. Thus, electron emitting locations are easier to detect and characterize by simple means \cite{19}. Another important advantage is that diamond appears to be a temperature sensitive material that has a nanodiamond to graphite transition at 1500--2000 K. One more key point has to be emphasized -- (N)UNCD is a two-phase material built of $sp^3$ diamond grains lined with $sp^2$ grain boundaries. Rich $sp^2$ content in (N)UNCD enables low turn-on fields $\sim1$ V/$\mu$m which in turn allows for large cathode-anode gaps between 100 to 200 microns, which is well enough to observe with ease processes that may evolve in the vacuum gap. By following this rationale, we were able to observe smooth and reversible transition from field emission to glow discharge formation in such nanodiamond vacuum DC diode. Our findings suggest that there is a cathodic plasma that forms due to thermal cathodic material runaway condition. Additional surface analysis with Raman spectroscopy revealed phase transformation from nanodiamond to nanographite that further confirms hot cathodic plasma scenario.

\begin{figure}
\includegraphics[scale=0.5]{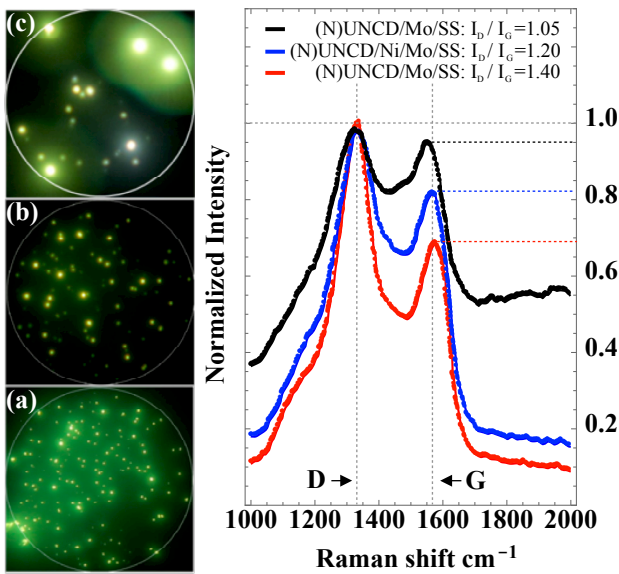}
\caption{A comparative chart between Raman spectra of $sp^2$-rich (N)UNCD and their laterally resolved emissivity. Electron emission micrographs (a), (b), (c) correspond to I$_D$/I$_G$ ratio 1.40 (red line), 1.20 (blue line) and 1.05 (black line), respectively.} 
\label{Fig:1}
\end{figure}

\begin{figure*}
\includegraphics[scale=0.67]{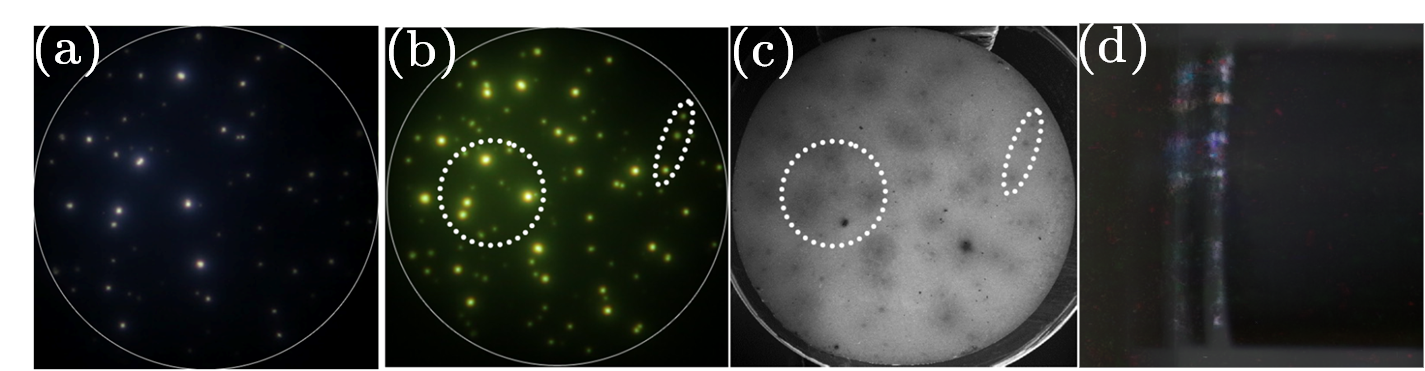}
\caption{(a-c) Field emission and scanning electron plan-view micrographs illustrating overlap between light and electron emission patterns and, in turn, overlap with the distribution of circular locations on (N)UNCD surface that were modified after the cathode performed for a few hours. Section (d) is a cross-sectional image of inter-electrode gap using optically polished stainless steel anode plate mirror illustrating that the blue/white light emission, section (a), can be related to micro-plasma formation near the cathode surface. The white circles in (a, b) depict the outer edge of the emitter stubs.} 
\label{Fig:2}
\end{figure*}

(N)UNCD films were grown by microwave-assisted chemical vapor deposition system in a mixture of CH$_4$/Ar/N$_2$ (with small addition of H$_2$ for initial plasma ignition) on stainless steel cylindrical stubs 4.4. mm in diameter using a standard procedure which was established in our previous studies \cite{6,19}.

The films were analyzed and assessed using a combination of three microscopy and spectroscopy methods. The relative graphitic $sp^2$ content in the diamond $sp^3$ matrix of the samples before and after field emission tests was evaluated by Raman spectroscopy via calculating the ratio between the intensities of the two major characteristic D and G peaks (labeled in Fig.\ref{Fig:1}), I$_D$/I$_G$ -- the smaller the ratio the larger the $sp^2$ content \cite{20}. A Renishaw (probe laser 633 nm) spectrometer and a Horiba (probe laser 532 nm) spectrometer were used. Scanning electron microscopy was used to inspect surface of the samples. Field emission properties of (N)UNCD were characterized and visualized using a field emission microscopy approach described in our previous work, see Refs.\cite{19, 21}. In such experiments, field emission from planar (N)UNCD surface is laterally resolved by making use of complementary pair of imaging anode screens. YAG:Ce/Mo screen (yttrium aluminum garnet crystal doped with cerium and coated with molybdenum) \cite{21} directly transfers the electron emission center distribution from the emitter surface through the gap, and ITO (semimetallic indium tin oxide)-on-glass screen collects and drains the field emitted electrons while passing light emitted from the field emitter surface. Such pair of imaging screens was originally introduced to look at CNT field emitters that are known to produce red light \cite{21}. By using YAG screen alone to evaluate emissivity of a series of (N)UNCD samples with varied microscopic and macroscopic roughness and the $sp^3$/$sp^2$ ratio \cite{19}, it was found that $sp^3$/$sp^2$ ratio (varied during synthesis) played the dominant role in boosting or lessening the output current which was consistent with previous findings \cite{22}. Even though emission was localized to a counted number of centers, the overall increase of the output current was related to the total emission area increase, from $\sim0.01\%$ to $\sim1\%$ of the entire cathode area (at maximum emission current) as I$_D$/I$_G$ ratio decreased from 1.6 to 1.2 suggesting an increase in $sp^2$ content from $3\%$ to over $6\%$. Along with the total emission area, the total output current increased by more than a factor of 10 and the turn-on electric field reduced from 11 V/$\mu$m to only 2.5 V/$\mu$m. This last finding was also consistent with previous results for spray deposited films comprised of diamond and graphite powders mixed at varied diamond-to-graphite ratios \cite{23}. Extending this correlation in order to further improve performance of (N)UNCD emitters, $sp^2$ content was further increased. This was accomplished by changing temperature in the reactor \cite{24}, while keeping the growth recipe the same as before \cite{6,19}. The resulting sample had I$_D$/I$_G$ ratio close to 1 suggesting the $sp^2$ content well over $10\%$. The summarized correlation between the emission area and the I$_D$/I$_G$ ratio is illustrated in Fig.\ref{Fig:1}. All the samples summarized in Fig.\ref{Fig:1} were $sp^2$ rich (I$_D$/I$_G\leq$1.4) as compared to canonical (N)UNCD that has I$_D$/I$_G$=1.60 \cite{25,19} when grown on refractory metal substrates such as Mo or W. The turn-on field was similar, close to 2.5 V/$\mu$m, and the total output current remained about the same at about the same maximal applied electric field. A big difference exists, though. It was resolved that the number of electron emission centers significantly decreased as the I$_D$/I$_G$ ratio decreased, from 1.40 to 1.20, to 1.05, while the apparent emission area around each center was getting larger as $sp^2$ content increased (I$_D$/I$_G$ decreased). Moving forward, two (N)UNCD samples with highest $sp^2$ content were compared and complementarily analyzed: see the electron emission micrograph \ref{Fig:1}b that corresponds to I$_D$/I$_G$=1.20 (blue Raman line) and the electron emission micrograph \ref{Fig:1}c that corresponds to I$_D$/I$_G$=1.05 (black Raman line).

Field emission from the sample Fig.\ref{Fig:1}b (I$_D$/I$_G$=1.20) was stable and reproducible during multiple ramp-up and ramp-down experiments. Therefore, this sample allowed for imaging with both YAG and ITO screens. Unlike CNT, (N)UNCD produced intense blue light. Most importantly, from the side by side comparison presented in Fig.\ref{Fig:2}, it is clear that the strongest electron emission locations imaged by YAG overlap precisely with blue light emission locations imaged by ITO. After a series of ramp-up/ramp-down experiments, the sample was examined $ex$ $situ$ by SEM. Extensive amount of circular dark spots was detected. The spot distribution across the surface, in turn, precisely overlapped with the YAG and ITO emission maps. To answer the question of the nature of the light generation, a few pictures were taken from a side view port between the cathode and anode, thanks to the large inter-electrode gap. The result shown in Fig.\ref{Fig:2}d was interpreted as an evidence for an early stage of plasma formation when plasma was confined to the cathode surface and laterally distributed around the strong electron emission locations, Fig.\ref{Fig:2}a and \ref{Fig:2}b.

\begin{figure*}
\includegraphics[scale=0.37]{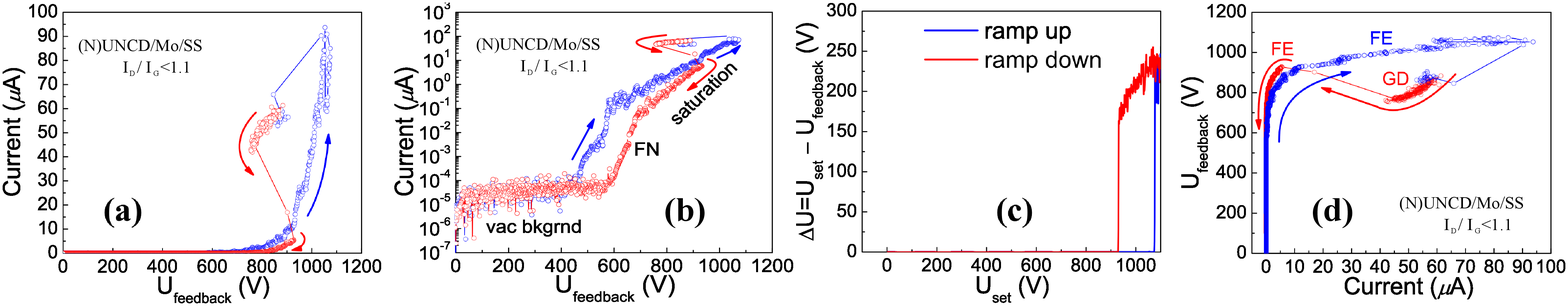}
\caption{Electric characteristics demonstrating three diode states plotted in linear (a) and semi-log (b) coordinates. The voltage loss plot is shown in the panel (c). Linear V-I representation emphasizing Fowler-Nordheim to glow discharge transition is plotted in the panel (d).} 
\label{Fig:3}
\end{figure*}

Compared to two other samples, the sample Fig.\ref{Fig:1}c (I$_D$/I$_G$=1.05) could possibly have segregation of the $sp^2$ phase resulting in that the number of electron emission centers significantly decreased but the apparent emission area around each center increased. Based on this observation, it was hypothesized that the current density per center could rise inflating the local heat load. To verify this hypothesis, the 1.05 emitter underwent the standard measurement protocol in which I-V curves were measured concurrently with front-view micrographs using the YAG screen. During the ramp-up, I-V curve retained a standard canonical Fowler-Nordheim form except for the saturation resistor-like behavior previously discussed in Refs.\cite{19,26}; see and compare linear and semi-log plots in Fig.\ref{Fig:3}a and \ref{Fig:3}b. Before the maximum voltage of 1100 V was reached (the limit voltage of the electrometer Keithley 2410 used in the experiments), and before the ramp-down started, the I-V curve switched into a new regime. The output current dropped and the actual feedback voltage dropped by over 200 V as compared to the pre-set value (pre-set voltage from the power supply). The voltage loss function is illustrated in Fig.\ref{Fig:3}c.

\begin{figure}[h]
\includegraphics[scale=0.7]{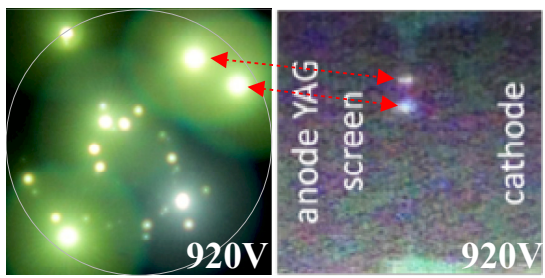}
\caption{Field emission front-view micrograph, illustrating the size and distribution of electron emission centers, placed side by side with side-view micrograph, illustrating two plasma conductors formation. The white circle depict the outer edge of the emitter stubs.} 
\label{Fig:4}
\end{figure}

The voltage loss coincided with the appearance of two plasma discharge channels that formed between and shortened the cathode and the anode screen. By comparing the front-view images and the cross section image combined in Fig.\ref{Fig:4}, it is clear that the two major centers dominating the electron emission (top right corner) were the two plasma channels seen inside the inter-electrode gap through the same side view port as in Fig.\ref{Fig:2}d. Going back to the I-V curve, the drastic change of the functional behavior can be correctly interpreted if the I-V is orthogonally rotated, i.e. re-plotted as a V-I curve. It is clear that the electrical characteristic of the vacuum diode undergoes two distinct transitions: 1) from a canonical FN to 2) a resistor-like behavior, to 3) a canonical glow discharge.

The same physical problem on FN to discharge transition is studied by Go and co-authors using atmospheric pressure Ar and air plasma environments\cite{27}. Remarkable difference exists though. The DC discharge of the vacuum diode, kept at base pressure $\sim$ 10$^{-9}$ Torr, is self-formed and self-stabilized. It is known that in order to stabilize a DC discharge inside a gas filled gap an external 100's of k$\Omega$ resistor should be connected between the diode and the external power supply \cite{27}. In our set up, the electrometer is directly connected to the diode. The built-in series (or ballast) resistance, a cause behind the current saturation effect (see the highlighted section on the semi-log I-V curve in Fig.\ref{Fig:3}b), comes into play during the discharge by stabilizing it. Starting in ultrahigh vacuum, the discharge is self-formed and self-supported. Summarizing the results, unlike in previous studies \cite{15,16}, our 
experimental evidence supports the cathodic plasma scenario. In this scenario, the cathode serves as a vehicle delivering atomized material for the discharge.

\begin{figure*}
\includegraphics[scale=0.49]{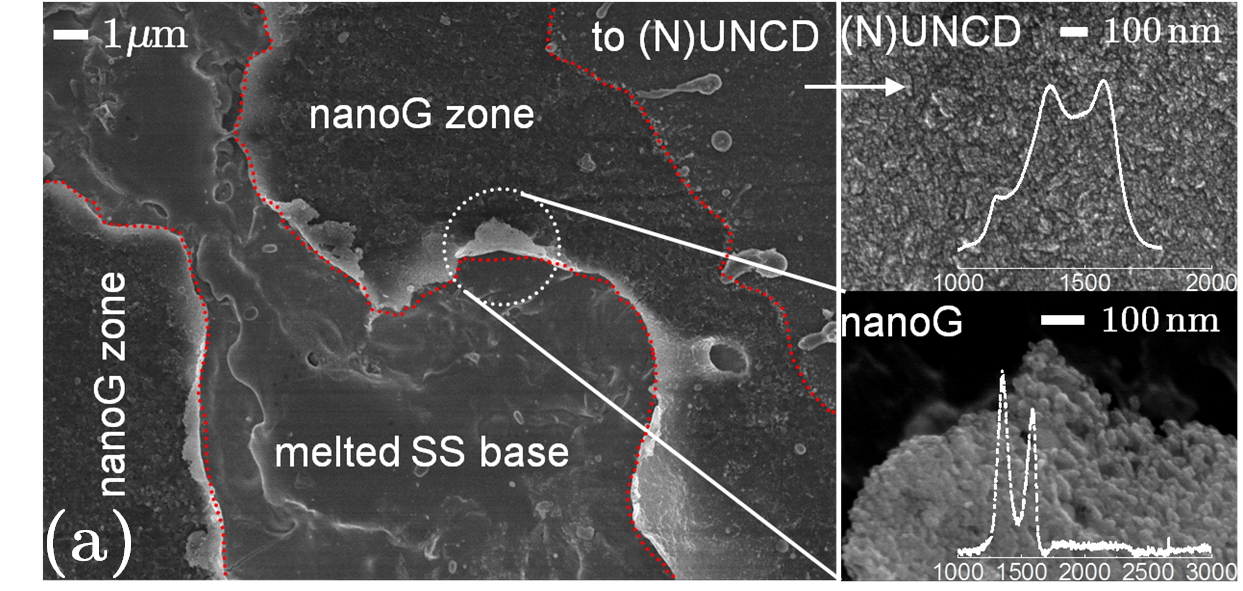}
\includegraphics[scale=0.57]{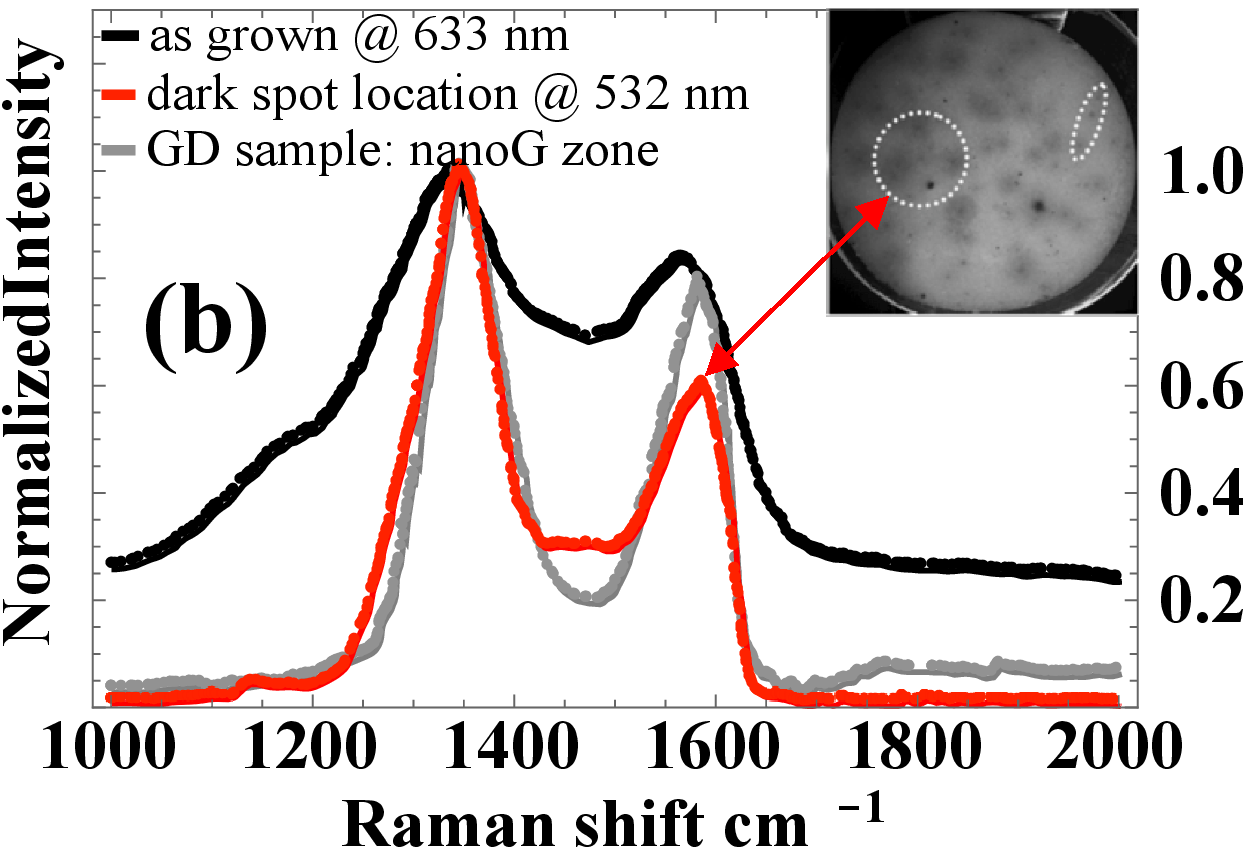}
\caption{Raman spectra for graphitic and UNCD sites on the damaged (a) and pre-damaged (b) cathodes due to electron field emission.} 
\label{Fig:5}
\end{figure*}

Having such evidence in hand, it is possible to estimate the local pressure in the gap using the Paschen's curve. Using the known gap, 100 $\mu$m, and discharge voltage, 1 kV, and vapor pressure diagram for carbon \cite{28}, the local pressure was estimated on the order of 10-100 Torr. Since atomized carbon discharges are not well studied, we rely on Paschen's curves known for noble gases or N$_2$ \cite{Pashen}. The estimated pressure of 10-100 Torr corresponds to a temperature of at least 3000 K.  Indirect confirmation of extensive heating of the cathode comes from post mortem analyses of the samples. Multiple damaged locations (localized breakdowns), revealed by SEM as shown in Fig.5a, consist of the re-melted metal substrate core surrounded by nanographite (labeled nanoG) which has a quite sharp transition into the rest of the cathode film that appears to be the pristine (N)UNCD film microscopically (SEM) and spectroscopically (Raman). We distinguish between nanodiamond and nanographite phases using micro-Raman. Diamond phase ($sp^3$) contained in (N)UNCD completely converted into nanographite. A few minute conversion time implies fast kinetics, and therefore suggest a temperature $\sim$2000K (see Chapter 13 in Ref.\cite{20}). Micro-Raman measurements made inside the circular, thermally pre-damaged, spots on the surface of the sample Fig.1b (I$_D$/I$_G$=1.20) captures a state of nanodiamond transitioning into nanographite state -- the recorded Raman spectrum suggests that the film state is neither of both but simultaneously contains highly separated nanographitic like D and G peaks (in the wrong ratio though) and nanodiamond $\omega_1$ peak near 1150 cm$^{-1}$. Further, probably time resolved experiments, would be needed to understand this non-trivial transition from nanodiamond to nanographite.

From analyzing the I-V curves in Fig.\ref{Fig:3}, the existence of visible light emission suggests thermal heating that may be dominated by Joule heating due to the diode-to-resistor transition. At the same time, the extremely high temperature regime capable of evaporating graphite and diamond has to have an initial stage before the Joule heating comes into play. We believe that the precursor rise in temperature can be explained by Nottingham process \cite{29}. According to analytical formulation, supported by the recent computational results \cite{11}, the inversion temperature should exist, that balances out the Joule effect and guarantees a stable emission within a certain I-$E$ domain for classical emitters based on tungsten and molybdenum tip arrays. Another recent work \cite{30} experimentally verified that the Nottingham process dominates in large area field emitters, such as CNT fibers, and therefore should play the most crucial role in setting their operating point temperature. The inversion temperature is contingent on the local electric field $E_{loc}$, work function $\phi$ and electron mass $m$ as \cite{29}
\begin{align}
T_i=A\frac{E_{loc}}{\sqrt{m \phi}}.
\end{align}
By replacing the free electron mass in this definition with effective electron mass, which was found to be as small as $1/18m_0$ from transport measurements\cite{33}, one can find that the inversion temperature is 4500 K which is close to the graphite vaporization/sublimation temperature. Thus Notingham heating, never inverting into cooling under our realistic conditions, and additional Joule heating drive the nanodiamond vacuum diode into thermal runaway regime. The use of the effective mass $1/18m_0$ relies on extensive experimental evidence that field emission and photoemission take places from graphitic like grain boundaries or graphitic patches \cite{23,31,32}. If this is the case, the effective electron mass in nanodiamond films is as small \cite{33} as it is in graphite \cite{34}.

 The observed plasma discharge stabilization for an extended period of time may be due to the fact that the surface temperature never reaches 4500 K. A somewhat lower temperature is expected due to a cooling mechanism when heavy atoms carry away significant amount of energy, i.e. cool the surface. Radiative cooling should also play a role, thou it contribution should be significantly less. Heats of vaporization $\Delta H$ for graphite and CNT, that are very similar $\approx$7 eV \cite{35}, and the Arrhenius formula relating evaporation rate $R$ and $\Delta H$ as $R\sim \exp(-\Delta H/kT)$ allows one to come up with a thermodynamical explanation of the remarkable lifetime difference between graphite-like and CNT emitters, i.e. enables a simple interpretation of why some materials are stable during field emission and some are not. Graphite like emitters heating to 3000--4000K are predicted to have a short lifetime and CNT emitters heating to 1000--2000 K are predicted to have a much more extended lifetime at much more extended output current range.

In conclusion, we have conducted field emission microscopy of two $sp^2$ rich ultrananocrystalline diamond films. We found that a glow plasma discharge can be triggered as the $sp^2$ content exceeds 10 at.\%. Put in short, electron emission self-induced heating through the Nottingham and Joule heating channels led to emitter material (carbon ions) evaporation. The evaporated plume was further ionized by co-existing flux of electrons and formed microplasma discharges that were stable for a certain amount of time enough to be observed and recorded. Such process could be stabilized by material evaporation that induces cooling. Our results are in agreement with previous black body light generation on CNT and extensive heat release measured for tungsten emitters \cite{29, 36}. Therefore, our findings support thermally driven cathodic plasma mechanism of the vacuum breakdown/arc. A pre-print from the KEK, a high energy physics institute in Japan, became available very recently \cite{37}. This work discusses intense light generation on the copper wall surfaces directly inside an operating large size 500 MHz megawatt RF cavity observed $in$ $situ$, in real time. Most breakdowns/arcs were connected with instabilities or appearance and then disappearance of the light emitting centers. Optical spectroscopy revealed that the emission is black body radiation, that temperatures in those locations are between 1500 and 2000 K and that the temperature increased with the electric field. This is a direct confirmation of the Nottingham effect. This additionally confirms the hot cathodic scenario of the vacuum arc formation. The results of this study obtained using copper RF structures are in agreement with our conclusions obtained using a DC nanodiamond vacuum diode. This means that detailed studies of the physics behind vacuum breakdown/arc taking place in large-scale high-power systems can be accomplished using alternative platforms that enable different diagnostics and therefore additional insights into the problem.

\begin{acknowledgments}
SSB was supported by the U.S. National Science Foundation under Award No. PHY-1549132, the Center for Bright Beams and under Award No. PHY-1535639. TN and SVB were supported by funding from the College of Engineering, Michigan State University, under
Global Impact Initiative.     
\end{acknowledgments}

\bibliography{Glow_discharge}

\end{document}